\documentclass[12pt]{amsart}

\begin{document}

\title[Global Medical Information System]
     {A Global Physician-Oriented Medical Information System}

\author{Axel Boldt}
\address{Department of Mathematics\\
Metropolitan State University\\
St. Paul, MN 55106, USA}
\email{Axel.Boldt@metrostate.edu}

\author{Michael Janich}
\address{B2 F5 Vantage Park\\
22 Conduit Rd\\
Mid Levels\\
Hong Kong}
\email{michael@janich.com}

\date{8 October 2008}

\begin{abstract}
We propose an Internet-based, free, world-wide, centralized
medical information system with two main target groups: practicing
physicians and medical researchers. After acquiring patients' consent,
physicians enter medical histories, physiological data and symptoms or
disorders into the system; an integrated expert system can then assist
in diagnosis and statistical software provides a list of the most
promising treatment options and medications, tailored to the
patient. Physicians later enter information about the outcomes of the
chosen treatments, data the system uses to optimize future treatment
recommendations. Medical researchers can analyze the aggregate data to
compare various drugs or treatments in defined patient populations on
a large scale.
\end{abstract}


\maketitle

\section{Introduction}

The two main tasks performed by physicians are: 
\begin{itemize}
\item diagnosing a disorder, based on presented symptoms, the
  patient's medical history and features, and ordered tests; and
\item choosing the most appropriate treatment or medication for a
      given disorder and a given patient.
\end{itemize}
In this article we propose a medical information system which aims to
assist physicians in both these tasks.

The ever increasing number of recognized diseases, combined with an
explosion in the number of marketed medications, poses formidable
challenges to the practicing physician. Many physicians rely mainly on
four information sources: their often outdated text books and lecture
notes, a small selection of medical journals, possibly biased
informational material from pharmaceutical companies, and their
personal or anecdotal experiences. All of these are less than ideal,
and we maintain that they can and should be supplemented by  more
rational decision aids based on modern data mining technology.

The core idea is simple: an expert system and treatment database that
physicians access over the Internet.  After acquiring informed
consent, they enter a patient's physiological data, symptoms and test
results and the expert system aids in diagnosis or recommends further
tests. Once the disorder is identified, the system recommends those
treatments or medications with the highest success probability for
that particular patient.  The physician chooses a treatment and later
records the outcome in the database. This outcome data is used by the
system to improve future treatment recommendations.

The system is designed to be used by physicians world wide, with
special regard for those working in developing countries. The use will
be free of charge and will only require an ordinary modem-speed
connection to the internet, something that is now available at
reasonable cost in most countries, often via mobile phones.

Since cost is often an important criterion when choosing a treatment
or medication, the system's recommendations will be accompanied by
cost estimates specific to the physician's location.

The two components of the system, diagnostic expert system and
treatment recommendation engine, use the same underlying patient
database but are logically independent and the system can go online as
soon as one of the two is fully functional. For example, physicians
can eschew the expert system altogether, rely on their own diagnostic
skills and directly ask for treatment recommendations for a particular
patient's disorder.

Apart from aiding physicians in their decision processes and thus
improving care and lowering cost on a global scale, the collected data
will also provide a rich resource for medical researchers. It will be
possible to easily determine the effectiveness of treatments and drugs
in various patient populations defined by combinations of
characteristics such as age, sex, ethnic group, pre-existing
conditions, lifestyle, or any of a large number of physiological
measurements recorded by the system. In addition, the system will be
immediately useful by helping to compare new, expensive and heavily
marketed medications with older, more established
alternatives. Evaluation of alternative treatments like acupuncture or herbal
remedies, heavily used all over the world but rarely studied in a
rigorous manner, will also become much easier. Newly emerging
epidemics will be detected in real time, much earlier than is possible today. 
Lastly, it is likely that mining the database will uncover
rare but severe side effects of established medications that have so
far escaped detection.

\section{Detailed Description}

\subsection{Recruitment of physicians}

The system will be open to all licensed practicing physicians
world-wide. Apart from the costs of a regular Internet connection, use
of the system will be free. Rather than attempting to contact all
physicians directly, or to advertise in relevant publications, it is
hoped that the national medical associations of the various countries
can be recruited to promote the system to their members and provide
them with authorized access codes. This approach has three advantages
over a more direct marketing campaign:

\begin{itemize}
\item it is significantly cheaper;
\item verification of physician's credentials is done by the
      organizations best suited for the task;
\item physicians extend a natural goodwill bonus to communications
      from their respective medical association.
\end{itemize}

\subsection{Physicians' interaction with the system}

Physicians are provided with access codes (passwords). They will
normally interact with the system by accessing the project's website
and logging in with their name and password.  The website will follow
W3-standards and carry few graphics in order to be easily accessable
through slow modem connections and simple devices, including mobile
devices. It will be designed to make the most common interactions
simple and fast.

When accessing the system for the first time, new users are directed
to a tutorial about the system's features and are required to agree to
a set of rules, mainly pertaining to patients' informed consent and
privacy (discussed below in section~\ref{privacysection}).

In the standard use case, after authentication physicians are
presented with a list of their patients and past treatments and are
invited to input outcome data for these treatments in a quick and
simple manner. They are then able
to view their patients' records, create new records, use the expert
system to receive diagnostic assistance, or use the statistical
database to get treatment/medication recommendations. The top-rated
treatments will be presented along with estimates of their success
probability and their cost. The system further facilitates the physician's decision
process by providing easy access to relevant background information,
such as as reviews from the Cochrane Library (\cite{cochrane}) and medical guidelines from
the clearinghouse responsible for the physician's location
(\cite{clear}, \cite{DeGeL}).

Signs, symptoms and diseases will be entered using the well-known
ICD-10 classification system (\cite{icd}), and information about
medications will be accessable through the standard Anatomical
Therapeutic Chemical Classification System (\cite{ATCCS}). It is to be
expected that these systems will not be entirely adequate for our
purposes and will have to be modified and extended to a certain
degree.

In addition to the direct web-based interface, the system will also
provide an XML-based interface, allowing for the easy data exchange
with physicians' other software applications, for instance with their
standard patient management software or with their systems of
interfacing with health insurers.

All patient information data transport will employ a layer of
encryption to avoid data being viewed or tampered with by unauthorized
third parties.

\subsection{Patient identifiers and privacy issues}
\label{privacysection}

The system will obey the world's strictest privacy laws, like those
commonly found in countries of the European Union. This requires in
particular that
\begin{itemize}
\item data is collected with full consent and can only be used for
  the express purposes given in the consent statement; and
\item patients retain the right to review their data and to have it
  removed at any time.
\end{itemize}

Physicians are required to obtain informed consent before they may
enter patients' data into the system. Physicians are not allowed to
reject treatment of patients who do not wish to participate in the system. 

No personally identifiable
information is ever entered into the database: no names, no exact
birthdates, no addresses etc. Instead, every patient is identified by
a PatientID, a simple number. The physician provides the patient with
their PatientID, so that the patient can authorize other physicians to
access their data if they so choose. Only physicians explicitly
authorized by a patient may access the records associated with that
patient's PatientID; physicians are required to keep proof of this
authorization on file. Physicians will typically keep a record of
all their patients' PatientIDs on their own computer.
 
It is possible and even to be expected that patients' names
together with corresponding PatientIDs will occasionally fall into the
wrong hands, for instance if malware is installed on a physician's
computer, their office is broken into or a
patient's private PatientID note is stolen. Since only physicians
associated with the PatientID may access the corresponding record, the
risk of data leaks is actually lower than in the current situation,
where all patient data would have been stored on the doctor's
compromised computer rather than in the online database.

Physicians, unlike patients, do not remain anonymous and are recorded
with full name and address in the database. Further, all interactions
of physicians with the system are logged. This allows to trace and
identify fraudulent use by non-physicians.

During scientific analysis the database will only be queried in the
aggregate and no individual record will be accessed in its
entirety. It is however important to note that this does not
completely protect against abuses; for instance a query of the form
"what percentage of adult black males living in Luxembourg and
standing less than 1.6 m tall are HIV positive", even though it uses
the database only in the aggregate, can still reveal very private
information about a small number of identifiable individuals. It is therefore
necessary that scientific analyses be reviewed ahead of time; see
section~\ref{sciencesection} below for more details on this process.

\subsection{Software}
\label{softwaresection}

To maximize transparency and to encourage others to submit bug fixes
and feature enhancements, the project will make use of existing Open
Source software whenever possible and all software written by the
project will be published under an Open Source license (\cite{opensource}).

Diagnostic expert systems have been developed before (\cite{hinf}; \cite{expert} for a comprehensive annotated list), and it is
unrealistic to expect the project to duplicate that work. Instead,
existing expert systems will be evaluated and the most appropriate one
will be licensed and extended. Most modern systems of this type employ
Bayesian networks, and these would benefit tremendously from the
statistical knowledge stored in the patient database.

The treatment recommendation engine will have to be written from
scratch, using existing algorithms that have been developed for data
mining applications in marketing; given a database containing customer
features and outcomes of past marketing strategies, these algorithms
can predict the most promising marketing strategy taylored to a given
customer. (See~\cite{datamining} for a survey of data mining
algorithms.) The algorithm will have to account for the fact that
information about patients is not necessarily complete; for example,
only some patient records will contain a recent blood glucose level.

As is usual in data mining applications, the comparatively low quality
of data collection procedures is compensated for by the large quantity
of data. A variety of different heuristics can be used to approach the
recommendation problem and a final algorithm can only be chosen after
empirical evaluation.

As mentioned above, the system requires information about the prices
of treatments and medications in the various countries. It is
possible and desirable that the collection and maintenance of this
data be carried out in collaboration with the various national medical
associations. The problem is non-trivial, since drug prices can vary
widely even within countries.

\subsection{Scientific analysis of the collected aggregate data}
\label{sciencesection}

For reasons outlined in section~\ref{privacysection} above, it is
necessary that all research proposals be reviewed ahead of
time. Credentialed medical researchers may submit research proposals
explaining the study's rationale, accompanied by the software that is
to analyze the database. After review of proposal and software, the
software is run against the live database and the results are returned
to the researcher. Research proposal, analysis software and results
are made public to ensure that negative results are not
suppressed. The entire process is free of charge for the researcher.

Toy systems with the same database structure as the live system (but
without real-world data) are provided to researchers, so that their
analysis software can be tested and debugged ahead of time. The use of
free software, as described in section~\ref{softwaresection} above,
will hopefully result in a rich ecosystem of analysis software freely
shared among researchers.

Every statistical analysis of the system's database needs to take into
account that the system does not and cannot guarantee that different
PatientIDs always correspond to different patients. Furthermore,
conclusions based on analysis of the database are of course valid only
for the subpopulation of patients who are cared for by participating
physicians and who have given consent to participating in
the system; this may not be a representative sample. External
studies comparing this subpopulation to the general patient population
will be highly desirable. 

As with all data mining applications,
correlations discovered in the data will have to be confirmed by
traditional randomized trials.

\subsection{Organizational structure and funding}

The system will be developed, deployed and run by a non-profit
organization, to be set up in a jurisdiction that grants tax-free
status to such organizations and that has strong privacy protection
laws. The organization will be assisted by a board of external
advisors.

To maintain neutrality and to avoid unduly influences on physicians'
decision processes, the system's website will not carry any
advertisings. The project will be financed completely by donations and
grants. These may come from individuals, companies (especially health
insurers), charity foundations, or national or international health
organizations. It is possible that an organization such as the NIH's
National Library of Medicine can be convinced to host and run the
system. An alternative model of funding, especially once the system is
established and accepted, would have the governments of
rich participating countries pay a (small) set amount per patient.

\section{Conclusion and Outlook}

The proposed system, once fully implemented, will improve world-wide
medical care in a number of important ways:
\begin{itemize}
\item practicing physicians, even in developing countries, will
      receive easy and free access to a medical expert system that can
      assist in diagnosis;
\item physicians receive promising treatment options, tailored to the
      patient, based on past collected outcome data;
\item inclusion of price data for treatments and medications allows
      physicians to choose the most cost-effective option in any given
      situation;
\item with patients' consent, medical histories stored in the database
      are easily transferrable from one physician to another;
\item creative use of the collected aggregate data will allow medical
      researchers to identify subpopulations of patients that respond
      particularly well to a certain medication or treatment;
\item comparisons of new drugs with established generic
      medications for the same condition become straightforward.
\end{itemize}

In the future, the system can be extended to incorporate
patients' genotype data, thereby representing an important step
towards the longstanding goal of truly personalized medicine.

In addition to these quite concrete and immediate benefits, we would
like to express our hope that widespread adoption of the system will
cause the profession of physician to evolve: from a passive container
of knowledge about symptom-disease correlations and disease-treatment
success probabilities to an active partner of the patient who
reassures, explains disorders and treatments, inquires and provides
advice about the patient's life circumstances and in general maximizes
the placebo effect in every way possible. This, we believe, will
ultimately turn out to be one of the main benefits of the proposed
system: the placebo effect, long considered a quirky nuisance by
western medicine, will return to its rightful place at the center of
the healer's work. Traditional systems of medicine will have much to
teach to physicians freed from the more mundane tasks of their
profession.


\begin{thebibliography}{9}

\bibitem{clear} G. Ollenschläger \emph{et al.} Improving the quality of health care: using
  international collaboration to inform guideline programmes by
  founding the Guidelines International Network (G-I-N). \emph{Quality
  \& safety in health care.} December 2004; 13(6):455-60.
\bibitem{cochrane} The Cochrane Collaboration. http://www.cochrane.org
  (accessed 27 September 2008).
\bibitem{icd} World Health Organization. \emph{The International
  Statistical Classification of Diseases and Health Related Problems
  ICD-10, Second Edition}, 2004
\bibitem{ATCCS} WHO Collaborating Centre for Drug Statistics
  Methodology. \emph{Anatomical Therapeutic Chemical Classification
  System.} 2008. \\
  http://www.whocc.no/atcddd/ (accessed 27 September 2008).
\bibitem{DeGeL} Hatsek A, Young O, Shalom E, Shahar Y. DeGeL: a
  clinical-guidelines library and automated guideline-support
  tools. \emph{Studies in health technology and informatics} 2008;139:203-12.
\bibitem{expert} Judith Federhofer. \emph{Medical Expert
  Systems. Doctor's Silent Partners.}\\
  http://www.computer.privateweb.at/judith/index.html (accessed 31
  December 2006, now only available through http://www.archive.org/).
\bibitem{opensource} 
  Andres M. St. Laurent. \emph{Understanding Open
  Source and Free Software Licensing.} O'Reilly Media, 2004.
\bibitem{hinf} E. Coiera. \emph{The Guide to Health Informatics (2nd Edition).} Arnold, London, October 2003. 
\bibitem{datamining} David Hand, Heikki Mannila, Padhraic
  Smyth. \emph{Principles of Data Mining.} MIT Press, 2001
\end{thebibliography}
\end{document}